\documentclass[proceedings, preprint]{rmaa}

\usepackage{paralist}

\usepackage{psfrag,color}

\SetYear{2024}
\SetConfTitle{III Workshop Astronomy for Inclusion}

\title{Astronomical images sonification: inclusion or outreach?} 

\author{
  J. Casado\altaffilmark{1,2} and B. García\altaffilmark{1,3}}

\altaffiltext{1}{Instituto en Tecnologías de Detección y Astropartículas (CNEA, CONICET, UNSAM), Mendoza, Argentina.}
\altaffiltext{2}{Instituto de Bioingeniería, Facultad de Ingeniería, Universidad de Mendoza, Argentina.}
\altaffiltext{3}{Universidad Tecnológica Nacional - Regional Mendoza, Argentina.}

\shortauthor{J. Casado and B. García}
\shorttitle{Image sonification}

\listofauthors{J. Casado and B. García}
\indexauthor{Casado, J.}
\indexauthor{García, B.}

\abstract{The field of sonification, using of non-speech audio for data analysis, is already established in space sciences. Meetings like “The audible Universe” focus on sonification tools applied in astronomy to represent complex data like nebulae and galaxies. Besides, little is said about the translation of images into sound, this challenge that seeks to represent data in 2 or 3 dimensions through a one-dimensional technique. The aforementioned leads to a total lack of consensus regarding the sound parameters to be used and how these are interpreted by people. This work seeks to delve deeper into the existing tools for image sonification, analyzing whether their objective is only outreach or includes the possibility of research. A new proposal is presented, maintaining sonoUno's software focus on research, pointing out the need for reliable techniques that integrate functional diversity people with an active role on research.}

\resumen{El campo de la sonorización, uso de sonido no hablado para el análisis de datos, se encuentra establecido dentro de las ciencias espaciales. ``The audible Universe'' es una convocatoria internacional que se centra en herramientas de sonorización aplicadas a la astronomía para representar datos complejos como nebulosas y galaxias. Sin embargo, poco se habla de la traducción de imágenes a sonido, este desafío que busca representar datos en 2 o 3 dimensiones a través de una técnica unidimensional. Lo mencionado lleva a que exista una total falta de consenso en cuanto a los parámetros de sonido a utilizar y cómo éstos son interpretados por las personas. Este trabajo busca profundizar sobre las herramientas para la sonorización de imágenes existentes, analizando si el objetivo de las mismas es solamente divulgación o incluye la posibilidad de investigación. Se presenta una nueva propuesta, manteniendo el enfoque del software sonoUno en investigación, puntualizando la necesidad de técnicas confiables que integren a las personas con diversidad funcional en el campo de la investigación.}

\addkeyword{astronomical data sonification}
\addkeyword{sonification of images}
\addkeyword{sonification: outreach}
\addkeyword{sonification: research}
\begin{document}
\maketitle

\section{Introduction}
\label{sec:intro}

Space sciences invest a significant amount of time and effort in the field of inclusion, and they have dedicated spaces in many of their annual conferences. From there, meetings have emerged about sonification, a starting science that proposes the use of sound for data analysis. Although the field of sonification has been around for 30 years since its first formal meetings and already three years of meetings dedicated to formalizing it in astronomy and astrophysics, much remains to be defined regarding how to represent complex image data, for example, a nebula, a galaxy, or a field of stars. Particularly, ``The Audible Universe'' is a workshop that has held two meetings (in 2021 and 2022) with specialists in sonification, focusing on tools applied in astronomy. It aims to establish a common framework and specific guidelines for the development of these resources \citep{ICADauduniv2023}. The experience of working in interdisciplinary groups, where both the data to be sonifyed and the target user of the tool were considered, is novel. The work method was similar to that used by the sonoUno team from the beginning, using tables and collaborative resources to visualize the same design/evaluation template, and even the need to generate training for everyone to learn to use sonification as a tool was highlighted.

Despite all these efforts, the field of image sonification for data analysis has more questions than answers. Transforming a three-dimensional space into sound is a massive challenge, there are some examples of tools that aim to solve this problem, but, no clear process has been described that allows a person to interpret an image solely through sonification. Most of the algorithms reduce three dimensions under two, the same technique used in sonoUno \citep{casadoJOSS2024}; particularly Chandra\footnote{Chandra sonification website: \url{https://chandra.si.edu/sound/#learnmore}} includes additional sound effects in the process to differentiate between celestial objects and produce a more pleasant sound.

\citet{unifiedterms2023} talk about unified terminology for sonification and visualization, they proposed to use an equivalent schema to the actual visualization theory, which adopts three basic constructs: marks (geometric objects representing the data, for example, a point or a line), channels (encode data attributes, for example, color), and spatial substrate (space where marks are positioned, for example, 1D cartesian plot). This author remarks the fact that sound is one-dimensional but recognizes that the human ears distinguish very well between frequencies and timbre. The authors propose the use of time as a spatial substrate and frequencies as marks, a congruent approach with the existing 1D sonification. However, nothing was mentioned about 2D or image sonification.

In this work, the current state of image sonification will be presented, evaluating whether the main goal of existing tools is outreach or if there are projects that seek to enable image analysis through sound, generating true inclusion.

\section{Image sonification existing tools}

As a starting field and with the lack of consensus, there are some scripts, projects with graphic user interfaces, and web-based developments for image sonification. Most of them convey the 2D image into a 1D array to sonify it. Image Sonification\footnote{Image sonification website: \url{https://sites.google.com/umich.edu/eecs351-project-sonify/home?authuser=0}}, VOSIS\footnote{VOSIS website: \url{https://vosis.app/}} and Photosounder\footnote{Photosounder website: \url{https://photosounder.com/}} are generic software to convey images into sound. They don't have a specific application, the motivation is to generate music from images. 

Regarding image sonification in astronomy and astrophysics, Chandra, Afterglow Access (AgA), and VoxMagelan are pioneers. About Chandra$^4$, the software is not open source, the sonification are made to make sense of each specific dataset, and they report work with the visually impaired community to test their results with good impact. The case of AgA software\footnote{Afterglow access website: \url{https://afterglow.skynetjuniorscholars.org/}} is a web-based tool that allows to open an image (.fits files) and sonify the light intensity by rows from bottom to top. This software was tested with screen readers and involved the visually impaired community during the development. Finally, VoxMagelan \citep{foran2022} was developed to sonify scatter plot images, allowing users to go through the images with the mouse pointer and sonify the point density; it also includes a visually impaired user during its development to ensure usability. All this project highlights the commitment of the astronomy and astrophysics communities to inclusion.

Besides, there exist some hands-on tools that allow the transformation of color into sound: a person has to move the electronic device through the image to be able to hear the sound, perceiving the spatial substrate with the hand position, leading to sound the color frequency information. These are the case of Orchestar\footnote{ Orchestar webpage: \url{https://astrolab.fas.harvard.edu/orchestar.html}} and Piano glove\footnote{Piano glove website: \url{https://learn.adafruit.com/pianoglove/what-youll-need}}, two open source tools that include instructions for costruction, wiring diagram and code in their websites.

\section{Image sonification with sonoUno}

The previously mentioned software are not open source, the only GitHub repository is for a library in C used by Photosounder. The ``sonoUno'' sonification software is a Python-based development \citep{casadoJOSS2024}. With their user centered design and during the user testing and exchange with other research groups, the need for image sonification arises.

The majority of image sonification computer tools produce sonification by reducing one row or column to only one value, representing the intensity value according to the complete image. The script presented here follows Equations \ref{eq:one} and \ref{eq:two} to obtain the sonification value from the image. Equation \ref{eq:one} represents the average value of the entire column to sonify. Then, equation \ref{eq:two} normalizes this value according to the maximum intensity of the image.

\begin{equation}
  \label{eq:one}
  x1 = \frac { \sum col\_values } { \sum col\_length }
\end{equation}

\begin{equation}
  \label{eq:two}
  to\_sonify = \frac { x1 } { max\_intensity }
\end{equation}

The script expects a path to locate the image at the moment of bash execution. Next, it shows a window with the image and waits for a key to start the sonification process. The scripts and documentation are available on GitHub\footnote{Image sonification GitHub: https://github.com/sonoUnoTeam/sonoUno-images}.

\begin{figure}[!t]\centering
  \includegraphics[width=0.8\columnwidth]{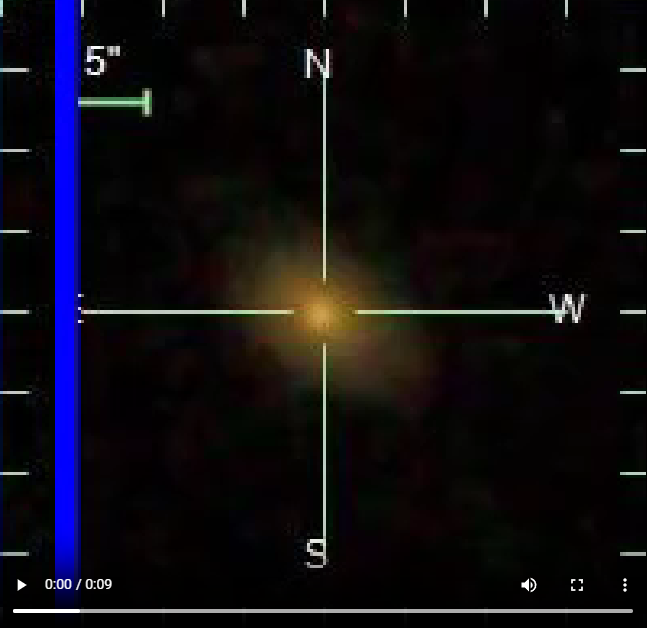}
  \caption{Screen capture during a galaxy image sonification with sonoUno for \textit{SDSS J115936.32-002841.8}.}
  \label{fig:one}
\end{figure}

Data from the SDSS database\footnote{SDSS dr16 database: \url{https://skyserver.sdss.org/dr16/en/home.aspx}} was used during the development process to test the script. Galaxy images were chosen because they represent light intensity with good contrast; consequently, the higher pitch corresponds to more light on the image, and a lower pitch represents less light. Examples with the image sonification can be recovered from the sonoUno website\footnote{SonoUno gallery: \url{https://www.sonouno.org.ar/galaxies-index/}}. Figure \ref{fig:one} shows the object \textit{SDSS J115936.32-002841.8}, the video\footnote{Galaxy YouTube video: \url{https://youtube.com/shorts/WWXARcFxZiQ?si=OqRnO4hU-bLk_xsa}} and spectrum sonification are in sonoUno gallery\footnote{Object sonification: \url{https://www.sonouno.org.ar/galaxy-sdss-j115936-32-002841-8/}}. 

In the video of the galaxy image sonification, there could be perceived noise coming from the marks and cardinal points in the image. That's because the algorithm sonify the image without any additional processing, ensuring the translation of the image directly to sound. For future work and after consulting with end-user needs, additional processing to convey those marks to auditory marks or better represent the 2D space, will be implemented.

\begin{figure}[!t]
  \makebox[0pt][l]{\textbf{a}}%
  \hspace*{0.5\columnwidth}
  \textbf{b}\\[-0.7\baselineskip]
  \parbox[t]{\columnwidth}{%
     \vspace{0pt}
     \includegraphics[width=0.5\columnwidth,height=3cm]{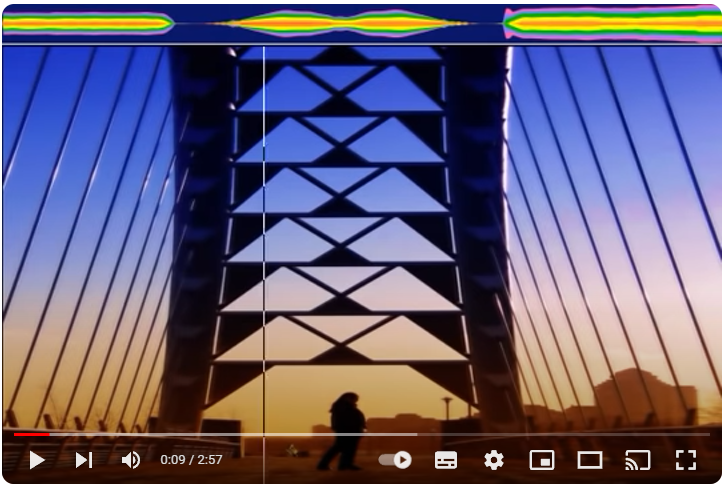}%
     \hfill%
     \includegraphics[width=0.5\columnwidth,height=3cm]{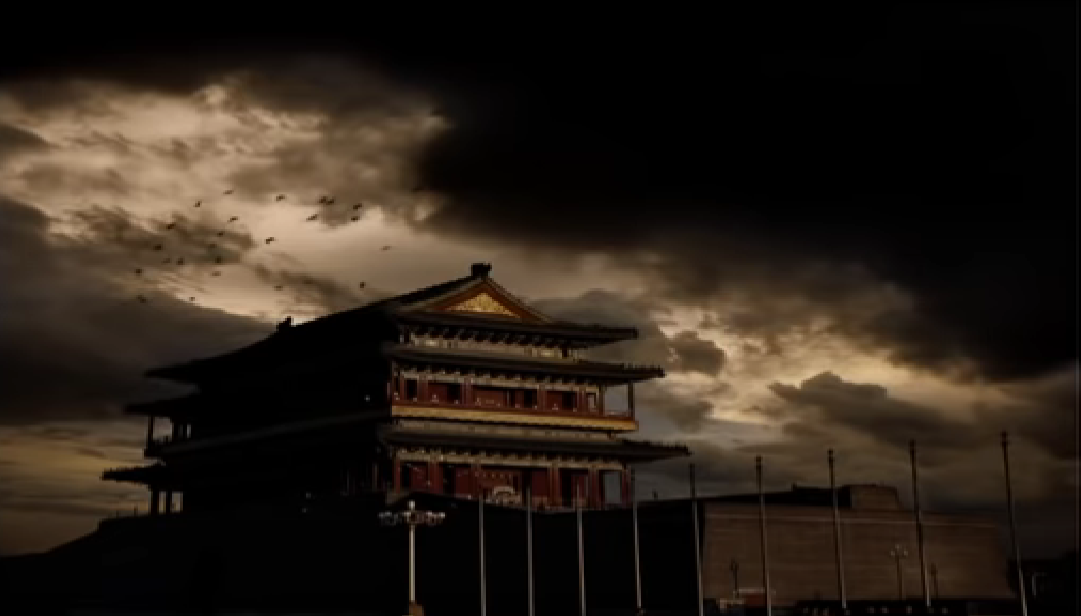}
     }
  

\caption{Images obtained from a Photosounder video to compare with the sonoUno sonification.}
  \label{fig:two}
\end{figure}

A comparison with Photosounder and AgA software was made. Photosounder represents a music-based tool; it offers a video demonstration on its website and a demo version; the other two software couldn't be downloaded (Image Sonification doesn't allow it, and the VOSIS free version only works on iPhone and iPad). AgA is an astronomical image sonification software; it was chosen because it is a free tool that allows image sonification and changes settings according to the user needs. Chandra only offers the videos, and the sonification presents musicalization. On the other hand, VoxMagelan was designed for scatter plots, not images.

Figure \ref{fig:two} contains two images obtained from Photosounder to be sonifyed with sonoUno. The images to compare with Photosounder were captured from a video in the main page of the project\footnote{\url{https://www.youtube.com/clip/UgkxrHEj6KmbuMSbC6nJRlfbPASYMV3cbzFN}}. The sonification of Figures \ref{fig:two}(\textit{a}) and \ref{fig:two}(\textit{b}) with sonoUno could be accessed on the sonoUno YouTube channel\footnote{https://youtu.be/zY3fgNhFiFA}. 

\begin{figure}[!t]\centering
  \includegraphics[width=\columnwidth]{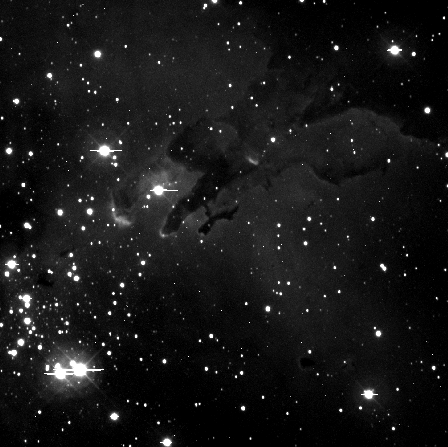}
  \caption{Image captured from AgA software, used to compare with the sonoUno sonification.}
  \label{fig:AgA}
\end{figure}

Concerning AgA software, the video with the sonification of Eagle Nebula (Figure \ref{fig:AgA}) with AgA  and sonoUno could be found in the sonoUno YouTube channel too\footnote{https://youtu.be/MnETEKBhJP8}. Note that AgA uses tick-marks to indicate the beginning and end of the sonification.

From these sonifications could be perceived a huge difference between sound, but essentially, the higher and lower pitches are similar. It's important to keep in mind that sonoUno maintains its developments far from musicalization because their main purpose is the use of sonification as a research tool. Besides, sonoUno paid attention to its users and their need for more pleasant sounds; these comparisons and new approaches are attempts to achieve a middle point that allows us to achieve a new feasible validation technique to be used in research.

\section{Discussion}

Most of the computer programs devoted to music are not open source, developed in complex languages, and not all of them present documentation. Different is the case with software related to astronomy or astrophysics, but they maintain the development close in some cases. Remarkably, all of the space science programs presented here take into consideration the visually impaired community.

Further, hands-on projects like Orchestar and Piano glove, allow one to perceive the spatial location inside the image through the position of the hand, a huge difference between this approach and some computer programs. Moreover, the major difference between hands-on projects and computer programs is the open-source concept of the first one. Not only the documentation is available, but also the diagrams, codes, and know-how to make that everything works properly. The same concept is maintained by sonoUno, which actual first approach to image sonification is available on GitHub$^{11}$.

Despite all these efforts, a clear process has not yet been described and validated that allows a person to interpret an image solely through sonification. During this III Workshop on Astronomy Beyond the Common Senses for Accessibility and Inclusion, \citet{IIIWAIzanella} presents a new approach using hand recognition in a webcam image to go through the image on the computer and sonify the point or area that the person touches. This project was initially intended for exhibitions and outreach, but the concept could be a good start to allow spatial positioning in image sonification. Following the actual state of sonification and the image sonification tools, the authors consider that a sonification algorithm that uses the mouse position and sonifys the color of an image (similar to VoxMagelan) could be a better approach to investigating how our brain perceives images through sound. Maybe using some of the open-source projects to start perception studies on image sonification techniques was the best choice to establish better sound parameters to perform image sonification. This is the only way to reach inclusion and makes possible the use of sonification in research fields.

\section{Conclusion}

At the moment, several projects are trying to make sense of images (three-dimensional objects) through sonification (one-dimensional space). However, a lack of consensus about how to translate images into sound is evident. The projects presented here don’t have documentation about perception studies, and some of them don't involve visually impaired people or how they understand the data behind images.

The developments are very promising, and the fact that functionally diverse people were involved in the projects establishes a remarkable precedent. Still, the exchange and use are made during outreach events without a framework devoted to validating the tool to enable research activities.

Based on the recommendations of the United Nations and relying on the ``Disability and Development Report''\footnote{Disability and Development Report: \url{https://www.un.org/development/desa/disabilities/wp-content/uploads/sites/15/2019/10/UN-flagship-report-on-disability-and-development.pdf}}, it is evident that long-term actions are needed to reinforce the fact that disability should not be a limiting factor for professional development.

Following \citet{IIIWAIvarano}, she uses the word justice during her talk and makes a difference with equality and equity; we must work to make environments accessible and not require tools to generate inclusion. Applying the concept to research, it would be ideal for the current way of working with scientific data to be accessible enough for anyone to do science, thanks to the multimodal approach.

\end{document}